\documentclass[prd,twocolumn,groupedaddress,showpacs,nofootinbib]{revtex4}
\usepackage{graphicx}
\usepackage{dcolumn}
\usepackage{bm}
\usepackage{amssymb}
\usepackage{mathrsfs}
\usepackage{amsmath}
\usepackage{epsfig}
\usepackage[dvips]{color}
\usepackage{hhline}

\begin{document}

\title{Resonant Leptogenesis and Verifiable Seesaw from Large Extra Dimensions }

\author{Pei-Hong Gu}
\email{peihong.gu@mpi-hd.mpg.de}

\affiliation{Max-Plank-Institut f\"{u}r Kernphysik, Saupfercheckweg
1, 69117 Heidelberg, Germany}

\begin{abstract}

In the presence of large extra dimensions, the fundamental scale
could be as low as a few TeV. This yields leptogenesis and seesaw at
a TeV scale. Phenomenologically two TeV-scale Majorana fermions with
a small mass split can realize a resonant leptogenesis whereas a
TeV-scale Higgs triplet with a small trilinear coupling to the
standard model Higgs doublet can give a verifiable seesaw. We
propose an interesting scenario where the small parameters for the
resonant leptogenesis and the type-II seesaw can be simultaneously
generated by the propagation of lepton number violation from distant
branes to our world.

\end{abstract}


\maketitle

In the theory of large extra dimensions \cite{add1998}, the Planck
scale of the 4-dimensional theory is related to that of the
$(4+n)$-dimensional theory by
\begin{eqnarray}
M_{\textrm{Pl}}^{2}\sim R^n_{}M_\ast^{n+2}\,,
\end{eqnarray}
where $M_{\textrm{Pl}}^{}\simeq 2.4\times 10^{18}_{}\,\textrm{GeV}$
is the reduced Planck mass and $R$ is the size of the extra
dimensions. Therefore the fundamental scale $M_\ast^{}$ of quantum
gravity could be as low as a few TeV for solving the hierarchy
problem between the electroweak and Planck scales. Clearly, all of
other scales in this theory couldn't be larger than a few TeV. This
can give interesting implications on neutrino physics. In this
direction, there have been many works
\cite{addm1998,ddg1998,ad1998,dk1999,mrs2000}. For example, ones
\cite{mrs2000} find the lepton number violation in a distant brane
can induce a very small trilinear coupling of the Higgs triplet to
the standard model (SM) Higgs doublet in our world. This naturally
makes the type-II \cite{mw1980} seesaw \cite{minkowski1977} model
accessible at colliders.

The low fundamental scale also constrains leptogenesis
\cite{fy1986,lpy1986,fps1995,pilaftsis1997,ms1998,pilaftsis1999,hs2004,di2006}
for baryon asymmetry. In the leptogenesis scenario, the decaying
particles should be very heavy for generating a sizable CP asymmetry
unless their masses are quasi-degenerate, which is simply input by
hand or is induced by radiative correction in some special models
\cite{fjn2003}, so that the CP asymmetry can be resonantly enhanced
\cite{fps1995,pilaftsis1997}. Remarkably the resonant leptogenesis
allows us to produce the baryon asymmetry below the low fundamental
scale. The author of \cite{pilaftsis1999} have studied the resonant
leptogenesis with bulk right-handed neutrinos. In the present work,
we shall introduce two right-handed neutrinos with equal but
opposite lepton numbers and localize them in our brane. In our
model, the lepton number is maximally broken in the distant branes
and then is shined to our brane by a bulk scalar. Therefore the
lepton number violation is highly suppressed in our world, where the
right-handed neutrinos mix together through a lepton number
conserving mass term and both have Yukawa couplings to the bulk
scalar so that they can naturally have a quasi-degenerate mass
spectrum. The induced Majorana fermions can accommodate a resonant
leptogenesis by their decays into the SM lepton and Higgs doublets.
Within this context, the small trilinear coupling of the Higgs
triplet to the SM Higgs doublet is also ready for the verifiable
type-II seesaw.

\vspace{2mm}

In our model, besides two right-handed neutrinos $N_{R_{1,2}^{}}^{}$
and the SM fields, there are three scalars: triplet $\xi$, doublet
$\eta$ and singlet $\chi$. We assign $L=1$ for $N_{R_{1}^{}}^{}$ and
$N_{R_{2}^{}}^{c}$ while $L=2$ for $\xi^\ast_{}$, $\eta$ and $\chi$.
The lepton number conserving interactions would be
\begin{eqnarray}
\label{lagrangian1} \mathcal{L}&\supset& -y_{\alpha
1}^{}\bar{\psi}_{L_\alpha^{}}^{}\phi N_{R_1^{}}^{}-y_{\alpha
2}^{}\bar{\psi}_{L_\alpha^{}}^{}\eta
N_{R_2^{}}^{}-M_N^{}\bar{N}_{R_1^{}}^{c}N_{R_2^{}}^{}\nonumber\\
&&-\frac{1}{2}h_1^{}\chi^\ast_{}\bar{N}_{R_1^{}}^{c}N_{R_1^{}}^{}-\frac{1}{2}h_2^{}\chi\bar{N}_{R_2^{}}^{c}N_{R_2^{}}^{}
-\rho\chi\eta^\dagger_{}\phi\nonumber\\
&&-\frac{1}{2}f_{\alpha\beta}^{}\bar{\psi}_{L_{\alpha}^{}}^c
i\tau_2^{}\xi \psi_{L_{\beta}^{}}^{} -\kappa_1^{}\chi^{}\phi^T_{}i
\tau_2^{}\xi \phi-\kappa_2^{}\chi^{\ast}\eta^T_{}i \tau_2^{}\xi
\eta\nonumber\\
&&+\textrm{H.c.}\,.
\end{eqnarray}
where $\psi_L^{}$ and $\phi$ denote the SM lepton and Higgs
doublets, respectively. The right-handed neutrinos
$N_{R_{1,2}^{}}^{}$, the triplet $\xi$, the doublet $\eta$ and the
SM fields are localized in our brane $(x,y^a_{}=0)$ while the
singlet $\chi$ propagates in the bulk. Here $a=1,...,n$ runs over
the extra dimensions. We further introduce a singlet scalar $\sigma$
with a lepton number $L=2$. This singlet is localized in a distant
brane $(x',y^a_{}=y^a_\ast)$. The transverse distance from the
distant brane to our brane is $r=|y_\ast^{}|$.

The singlet $\chi$ can interact with our brane through its couplings
to the right-handed neutrinos and the triplet and doublet scalars as
shown in Eq. (\ref{lagrangian1}). In the distant brane, it can also
have lepton number conserving interaction with the singlet $\sigma$
\cite{addm1998},
\begin{eqnarray}
\label{lagrangian2} L&\supset&\int d^4_{}x'
M_\ast^2\sigma(x',y^a_\ast)\chi^\ast_{}(x',y^a_\ast)\,.
\end{eqnarray}
So, the singlet $\chi$ can mediate the communication from the
distant brane to ours. In particular, it can carry the lepton number
violation to our world through its shined value
$\langle\chi\rangle$,
\begin{eqnarray}
\langle\chi(x,y^a_{}=0)\rangle=\langle\sigma(x,y=y^a_\ast)\rangle\Delta_n^{}(r)\,.
\end{eqnarray}
Here the vacuum expectation value (VEV) $\langle\sigma\rangle$ acts
as a point source whereas $\Delta_n^{}(r)$ is the Yukawa potential
in the $n$ transverse dimensions \cite{ad1998,mrs2000},
\begin{eqnarray}
\Delta_n^{}(r)=\frac{1}{(2\pi)^{\frac{n}{2}}_{}}\left(\frac{m_\chi^{}}{M_\ast^{}}\right)^{n-2}_{}\left(m_\chi^{}
r\right)^{-\frac{n-2}{2}}_{}K_{\frac{n-2}{2}}^{}\left(m_\chi^{}
r\right)\,.
\end{eqnarray}
With the natural choice $r=R$ and $\langle\sigma\rangle \lesssim
M_\ast^{}$, it is easy to read
\begin{eqnarray}
\langle\chi\rangle\sim M_\ast^{}\Delta_n^{}(R)\,.
\end{eqnarray}
We will clarify later a heavy mass $m_\chi^{}$ is necessary for a
successful leptogenesis. However, the modified Bessel function
$K_{\frac{n-2}{2}}^{}\left(m_\chi^{} r\right)$ will exponentially
suppress $\langle\chi\rangle$ if $m_\chi^{}\gg 1/r$. This
$\langle\chi\rangle$ is too small to generate the desired neutrino
masses. Therefore, we consider the brane-lattice crystallization
scenario \cite{adm1998} where the bulk is populated with large
numbers of branes. One finds the lepton number violation in our
brane would be \cite{addm1998}
\begin{eqnarray}
\langle\chi\rangle&\sim&M_{\ast}^{}\int d^n_{}r
n_{brane}^{}\Delta_n^{}(r)\nonumber\\
&=&M_\ast^{}n_{brane}^{}\left(\frac{m_\chi^{}}{M_\ast^{}}\right)^{n-2}_{}\frac{1}{m_\chi^n}\nonumber\\
&=&M_\ast^{}\left(\frac{M_\ast^{}}{m_\chi^{}}\right)^{2}_{}\left(\frac{M_\ast^{}}{M_{\textrm{Pl}}^{}}\right)^{\frac{4}{n}}_{}
\end{eqnarray}
with the brane density \cite{adm1998}
\begin{eqnarray}
n_{brane}^{}\sim M_\ast^n
\left(\frac{M_\ast^{}}{M_{\textrm{Pl}}^{}}\right)^{\frac{4}{n}}_{}\,.
\end{eqnarray}
By taking the natural assumption $m_{\chi}^{}\lesssim M_\ast^{}$,
the VEV $\langle\chi\rangle$ is power suppressed by the ratio of
$M_\ast^{}$ over $M_{\textrm{Pl}}^{}$. In the following we will
consider
\begin{eqnarray}
\label{input1} \langle\chi\rangle\sim
260\,\textrm{eV}~~\textrm{for}~n=6\,,M_\ast^{}=3\,\textrm{TeV}~
\textrm{and}~m_\chi^{}\lesssim M_\ast^{}\,.
\end{eqnarray}

\vspace{2mm}

Due to the VEV $\langle\chi\rangle$, the masses of the right-handed
neutrinos $N_{R_{1,2}^{}}^{}$ would be
\begin{eqnarray}
\label{lagrangian3}
\mathcal{L}&\supset&-M_N^{}\bar{N}_{R_1^{}}^{c}N_{R_2^{}}^{}-\frac{1}{2}m_1^{}\bar{N}_{R_1^{}}^{c}N_{R_1^{}}^{}
-\frac{1}{2}m_2^{}\bar{N}_{R_2^{}}^{c}N_{R_2^{}}^{}\nonumber\\
&& +\textrm{H.c.}
\end{eqnarray}
with
\begin{eqnarray}
\label{mass1} m_{1,2}^{}=h_{1,2}^{}\langle\chi\rangle\ll M_N^{}\,.
\end{eqnarray}
We can diagonalize the above mass terms to be
\begin{eqnarray}
\mathcal{L}&\supset& -\frac{1}{2}M_{\pm}^{}\bar{X}_R^{\pm
c}X_{R}^{\pm}+\textrm{H.c.}
\end{eqnarray}
by taking the rotations as below,
\begin{subequations}
\begin{eqnarray}
N_{R_1^{}}^{}&=& c X_{R}^{+}-is X_{R}^{-}\,,\\
N_{R_2^{}}^{}&=& s X_{R}^{+}+i c X_{R}^{-}\,.
\end{eqnarray}
\end{subequations}
Here we have take the following notations,
\begin{eqnarray}
\label{mixing1} c\equiv \cos \vartheta\,,~~s\equiv \sin \vartheta
~~\textrm{for}~~\vartheta =\frac{1}{2}\arctan\frac{2M_N^{}
}{m_2^{}-m_1^{}}
\end{eqnarray}
and
\begin{subequations}
\label{mass2}
\begin{eqnarray}
M_{+}^{}&=&2scM_N^{}+c^2_{}m_1^{}+s^2_{}m_2^{}\,,\\
M_{-}^{}&=&2scM_N^{}-s^2_{}m_1^{}-c^2_{}m_2^{}\,.
\end{eqnarray}
\end{subequations}
Without loss of generality we will assume $m_{1}^{}< m_{2}^{}$ so
that
\begin{eqnarray}
M_+^{}>M_-^{}>0\,.
\end{eqnarray}
Actually, the small and large masses (\ref{mass1}), i.e.
$m_{1,2}^{}\ll M_N^{}$ will induce
\begin{subequations}
\begin{eqnarray}
\label{mixing2}
&\vartheta\simeq \frac{\pi}{4}\,,&\\
\label{mass3} &M_\pm^{}\simeq M_N^{}\pm\frac{1}{2}(m_1^{}+m_2^{})\gg
M_+^{}-M_-^{}\,.&
\end{eqnarray}
\end{subequations}
It is convenient to define the Majorana fermions
\begin{subequations}
\begin{eqnarray}
X_{}^{+} &=& X_{R}^{+}+X_{R}^{+c}\,,\\
X_{}^{-} &=& X_{R}^{-}+X_{R}^{-c}\,,
\end{eqnarray}
\end{subequations}
as the mass eigenstates,
\begin{eqnarray}
\mathcal{L}&\supset& -\frac{1}{2}M_{\pm}^{}\bar{X}_{}^{\pm
}X_{}^{\pm}\,.
\end{eqnarray}
We further assume the doublet scalar $\eta$ much heavier than the
Majorana fermions $X^\pm_{}$, say $m_\eta^2\gg M_{\pm}^2$. In this
case, we can integrate out $\eta$ and then simplify the Yukawa
couplings given by the first line of Eq. (\ref{lagrangian1}) in a
new form,
\begin{eqnarray}
\label{lagrangian3} \mathcal{L}&\supset& -\left(cy_{\alpha
1}^{}-sy'^{}_{\alpha 2}\right)\bar{\psi}_{L_\alpha^{}}^{}\phi
X_{}^{+}\nonumber\\
&&-i\left(sy_{\alpha 1}^{}+cy'^{}_{\alpha
2}\right)\bar{\psi}_{L_\alpha^{}}^{}\phi X_{}^{-}+\textrm{H.c.}
\end{eqnarray}
with
\begin{eqnarray}
\label{yukawa} y'^{}_{\alpha 2}=-y^{}_{\alpha
2}\frac{\rho\langle\chi\rangle}{m_\eta^2}\,.
\end{eqnarray}

For $M_\pm^{}\gg M_+^{}-M_-^{}$, we can follow the standard method
\cite{pilaftsis1997} of the resonant leptogenesis \footnote{The
right-handed neutrinos for the type-I seesaw will induce the vertex
loop besides the self-energy loop in their decays. At the same time,
the Higgs triplet for the type-II seesaw, which will be introduced
later, will mediate another vertex correction in the decays of the
right-handed neutrinos. The right-handed neutrinos and the Higgs
triplet are at the TeV scale so that the CP asymmetry constrained by
the neutrino masses would be too small unless we take the resonant
enhancement into account. However, the resonant effect only exists
in the self-energy loop. So, the vertex corrections induced by the
Higgs triplet or the right-handed neutrinos would have no
significant contributions to the leptogenesis.} to compute the
lepton asymmetry from the decays of per $X_{}^\pm$,
\begin{eqnarray}
\label{cpasymmetry} \varepsilon_{X_{}^\pm}^{}
&=&\frac{\Sigma_{\alpha}^{}\left[\Gamma(X_{}^\pm \rightarrow
\psi_{L_\alpha^{}}^{} + \phi_{}^{\ast})-\Gamma(X_{}^\pm \rightarrow
\psi_{L_\alpha^{}}^{c} +
\phi_{}^{})\right]}{\Sigma_{\alpha}^{}\left[\Gamma(X_{}^\pm
\rightarrow \psi_{L_\alpha^{}}^{} + \phi_{}^{\ast})+\Gamma(X_{}^\pm
\rightarrow \psi_{L_\alpha^{}}^{c} +
\phi_{}^{})\right]}\nonumber\\
\vspace{10mm}
&\simeq&\frac{sc\Sigma_{\alpha\beta}^{}\left(|y_{\alpha
1}^{}|^2_{}-|y'^{}_{\alpha 2}|^2_{}\right)} {4\pi
A_{X_{}^\pm}^{}}\nonumber\\
&&\times\textrm{Im}\left(c^2_{}y_{\beta 1}^{\ast}y'^{}_{\beta
2}-s^2_{}y'^{\ast}_{\beta 2}y^{}_{\beta 1}\right)\frac{x}{x_{}^{2}
+\frac{1}{64\pi^2_{}}A_{X_{}^\mp}^{2}}
\end{eqnarray}
with
\begin{subequations}
\begin{eqnarray}
&A_{X_{}^+}^{}=\Sigma_{\alpha}^{}|cy_{\alpha 1}^{}-sy'{}_{\alpha
2}|^2_{}\,,&\\
&A_{X_{}^-}^{}=\Sigma_{\alpha}^{}|sy_{\alpha
1}^{}+cy'{}_{\alpha 2}|^2_{}\,,&\\
\label{ratio} &x=\frac{M_+^2-M_-^2}{M_{+}^{}M_{-}^{}}\,.&
\end{eqnarray}
\end{subequations}
In the weak washout region, i.e.
\begin{eqnarray}
\Gamma_{X_{}^\pm}^{}<H(T)\left|_{T\simeq M_{\pm}^{}}^{}\right.\,,
\end{eqnarray}
the final baryon asymmetry can be approximately given by
\cite{kt1990}
\begin{eqnarray}
\frac{n_B^{}}{s} &=& \frac{28}{79} \frac{n_{B}^{}-n_{L}^{}}{s} =
-\frac{28}{79}\frac{n_{L}^{}}{s}
\nonumber\\[3mm]
&\simeq& -\left.
\frac{28}{79}\varepsilon_{X^\pm_{}}^{}\frac{n_{X_{}^{\pm}}^{eq}}{s}\right|_{T\simeq
M_{\pm}^{}}^{} \nonumber\\
&\simeq&
-\frac{1}{\mathcal{O}(10)}\frac{\varepsilon_{X^\pm_{}}^{}}{g_\ast^{}}\,.
\end{eqnarray}
Here $\Gamma_{X_{}^\pm}^{}$ is the decay width
\begin{eqnarray}
\Gamma_{X_{}^\pm}^{} &=&\frac{1}{8\pi}A_{X_{}^\pm}^{}M_{\pm}^{}\,,
\end{eqnarray}
whereas $H(T)$ is the Hubble constant
\begin{eqnarray}
H(T)&=&\left(\frac{\pi^{2}_{}g_{\ast}^{}}{90}\right)^{\frac{1}{2}}_{}
\frac{T^{2}_{}}{M_{\textrm{Pl}}^{}}
\end{eqnarray}
with $g_{\ast}^{}\simeq 106.75$ being the relativistic degrees of
freedom.

We should keep in mind that the right-handed neutrinos
$N_{R_{1,2}^{}}^{}$, or equivalently the Majorana fermions
$X^\pm_{}$ have Yukawa couplings with the bulk scalar $\chi$. The
induced annihilations of $X^\pm_{}$ should go out of equilibrium.
This can be achieved if $X^\pm_{}$ is much lighter than the bulk
scalar $\chi$ and the triplet scalar $\xi$. Actually, a factor
$\displaystyle{m_{\chi}^{}/M_{\pm}^{}}\sim 3-10$ is enough for the
decoupling of the t-channel processes
$N_{R_{1,2}^{}}^{c}N_{R_{1,2}^{}}^{}\rightarrow \chi^\ast_{}\chi$
which is fast Boltzmann suppressed at temperatures below the mass
$m_\chi^{}$. In addition, the annihilations of $N_{R_{1,2}^{}}^{}$
to the SM Higgs doublet $\phi$ (through the s-channel exchange of
$\chi$) is highly suppressed by the shined value
$\langle\chi\rangle$. Furthermore, the coupling of $\chi$ to $\xi$
and $\phi$ will also lead to annihilations of $N_{R_{1,2}^{}}^{}$.
Such processes can be suppressed if $\xi$ is much heavier than
$X^\pm_{}$, say $\displaystyle{m_{\xi}^{}/M_{\pm}^{}} \sim 3-10$. As
for the lepton number violating processes mediated by $\xi$, they
will not wash out the produced lepton asymmetry due to the small
$\langle\chi\rangle$. As a result of the symmetry breaking of the
global lepton number, there would be a massless Goldstone ---
majoron \cite{cmp1980}, which is composed of the imaginary parts of
the triplet $\xi$ and the doublet $\eta$ in our brane, the singlet
$\chi$ in the bulk and the singlets $\sigma$ in the distant branes.
Because $\xi$, $\eta$ and $\chi$ with small VEVs only contribute a
tiny fraction, this majoron is harmless \cite{mrs2001}, in
particular, it will not significantly affect the annihilation of
$N_{R_{1,2}^{}}^{}$ \cite{gs2009}.

We now give a numerical estimation. With the input (\ref{input1}),
we can obtain
\begin{eqnarray}
\label{input2} M_\pm^{}\simeq 600\,\textrm{GeV} ~~\textrm{and}~~x=
\mathcal{O}(10^{-12}_{})
\end{eqnarray}
by inserting
\begin{eqnarray}
\label{input3} M_N^{}=600\,\textrm{GeV}
~~\textrm{and}~~h_{1,2}^{}\sim \mathcal{O}(10^{-3}_{})
\end{eqnarray}
to Eqs. (\ref{mass1}), (\ref{mixing1}), (\ref{mass2}) and
(\ref{ratio}). We further consider
\begin{eqnarray}
\label{input4} \rho=m_\eta^{}\lesssim M_\ast^{}
\end{eqnarray}
in Eq. (\ref{yukawa}) and then get
\begin{eqnarray}
\label{input5} y'^{}_{\alpha 2} \sim \mathcal{O}(10^{-10}_{})\quad
\textrm{for}\quad y^{}_{\alpha 2} \sim \mathcal{O}(1)\,.
\end{eqnarray}
We also take
\begin{eqnarray}
\label{input6} |y_{\alpha 1}^{}|= \mathcal{O}(10^{-7}_{}) ~~
\textrm{and} ~~\sin\frac{y_{\alpha 1}^{}}{|y_{\alpha
1}^{}|}=\mathcal{O}(0.1)\,.
\end{eqnarray}
With the parameter choice (\ref{input2}), (\ref{input5}) and
(\ref{input6}), it is easy to derive the CP asymmetry
\begin{eqnarray}
\varepsilon_{\pm}^{}=\mathcal{O}(10^{-7}_{})
\end{eqnarray}
and hence the baryon asymmetry
\begin{eqnarray}
\frac{n_B^{}}{s}=\mathcal{O}(10^{-10}_{})\,.
\end{eqnarray}

\vspace{2mm}

The left-handed neutrinos can obtain a Majorana mass matrix with two
nonzero eigenvalues from the two right-handed neutrinos
$N_{R_{1,2}^{}}^{}$ as a result of inverse
\cite{mohapatra1986,mv1986} seesaw. However, these masses are too
tiny to explain the observed neutrino oscillations because of the
smallness of the Yukawa couplings $y_{\alpha 1}^{}$ and
$y'^{}_{\alpha 2}$. Alternatively, our model accommodates the
type-II seesaw,
\begin{eqnarray}
\label{lagrangian4} \mathcal{L}&\supset&
-\frac{1}{2}f_{\alpha\beta}^{}\bar{\psi}_{L_{\alpha}^{}}^c
i\tau_2^{}\xi \psi_{L_{\beta}^{}}^{} -\mu\phi^T_{}i \tau_2^{}\xi
\phi+\textrm{H.c.}\,.
\end{eqnarray}
Here the trilinear coupling $\mu$ of the Higgs triplet $\xi$ to the
SM Higgs doublet $\phi$ is given by the shined value of the bulk
field $\chi$,
\begin{eqnarray}
\mu=\left(\kappa_1^{}+\kappa_2^{}\frac{\rho^2_{}\langle\chi\rangle^2_{}}{m_\eta^4}\right)\langle\chi\rangle\,.
\end{eqnarray}
With the previous parameter choice for the leptogenesis, the
neutrino masses would be
\begin{eqnarray}
(m_\nu^{})_{\alpha\beta}^{}&=&f_{\alpha\beta}^{}\langle\xi\rangle
\simeq-f_{\alpha\beta}^{}\frac{\mu\langle\phi\rangle^2_{}}{m_\xi^2}\nonumber\\
&\simeq&-f_{\alpha\beta}^{}\kappa_1^{}M_\ast^{}\frac{\langle\phi\rangle^2_{}}{m_\xi^{2}}\left(\frac{M_\ast^{}}{m_\chi^{}}\right)^{2}_{}\left(\frac{M_\ast^{}}{M_{\textrm{Pl}}^{}}\right)^{\frac{4}{n}}_{}\nonumber\\
&=&-f_{\alpha\beta}^{}\times \frac{\kappa_1^{}}{0.1} \times
\left(\frac{M_\ast^{}}{m_\xi^{}}\right)^{2}_{}\times
0.086\,\textrm{eV}\,.
\end{eqnarray}
For a natural choice
\begin{eqnarray}
m_\xi^{}\lesssim M_\ast^{}\,,
\end{eqnarray}
the Yukawa couplings $f_{\alpha\beta}^{}$ of the Higgs triplet $\xi$
to the left-handed lepton doublets $\psi_{L_{\alpha\beta}^{}}^{}$
should be sizable to give the expected neutrino masses.

\vspace{2mm}

Since the Higgs triplet $\xi$ with the mass $m_\xi^{}\lesssim
M_\ast^{}$ is kinematically accessible at the LHC and at future
colliders whereas its sizable Yukawa couplings $f_{\alpha\beta}^{}$
to the SM lepton doublets $\psi_{L_{\alpha,\beta}^{}}^{}$ determine
the texture of the neutrino mass matrix, the neutrino masses can be
verified \cite{krr2007,ghsz2009} by the decays of $\xi^{\pm\pm}$
into the charged leptons
$l_{L_\alpha^{}}^{\pm}l_{L_\beta^{}}^{\pm}$. At the same time, the
doublet $\eta$ with the mass $m_\eta^{}\lesssim M_\ast^{}$ has a
Yukawa coupling $y_{\alpha 2}^{}\sim \mathcal{O}(1)$ to the SM
lepton doublet $\psi_{L_{\alpha}^{}}^{}$ and the right-handed
neutrino $N_{R_2^{}}^{}$. Through the detection on the decays
$\eta^{-}_{}\rightarrow l_{L_\alpha^{}}^{-}N_{R_2^{}}^c$ and
$\eta^{+}_{}\rightarrow l_{L_\alpha^{}}^{+}N_{R_2^{}}^{}$ and/or on
the annihilations $l_{L_\alpha^{}}^{+}l_{L_\beta^{}}^{-}\rightarrow
N_{R_2^{}}^{c}N_{R_2^{}}^{}$, the right-handed neutrino
$N_{R_2^{}}^{}$ and then the Majorana fermions $X^\pm_{}$ could be
found as a missing energy.

\vspace{2mm}

In summary, the theory with the large extra dimensions implies a low
fundamental scale of the order of TeV. In this scenario the resonant
leptogenesis becomes attractive as it can generate the baryon
asymmetry at a TeV scale. The resonant leptogenesis requires a tiny
mass difference between the decaying particles. The low fundamental
scale also constrains the Higgs triplet for the type-II seesaw at
the TeV scale. The neutrino masses can be verified in presence of a
very small trilinear coupling between the triplet and doublet Higgs
scalars. We show the small parameters for the resonant leptogenesis
and the verifiable type-II seesaw can be simultaneously achieved by
the shined lepton number violation from the distant branes to our
world. In our model, it is possible to detect the existence of the
decaying Majorana fermions for the resonant leptogenesis, besides
the neutrino masses from the type-II seesaw with the Higgs triplet.

\vspace{2mm}

\textbf{Acknowledgement}:  I thank Manfred Lindner for hospitality
at Max-Plank-Institut f\"{u}r Kernphysik. This work is supported by
the Alexander von Humboldt Foundation.

\end{document}